\documentclass[aps,prl,twocolumn,showpacs,superscriptaddress,groupedaddress]{revtex4}  
\usepackage{graphicx}  
\usepackage{dcolumn}   
\usepackage{bm}        
\usepackage{xspace,amssymb,mathdots}   
\usepackage{amsmath,mathrsfs,
color,mathdots, bbold,tikz, xspace,version}
\usepackage{dsfont}
\def\Sbb{{\mathbb S}}
\def\Nbb{{\mathbb N}}
\def\Id{{\mathbb 1}}

\def\xbf{{\mathbf x}}
\def\fbf{{\mathbf f}}
\def\psibf{{\mathbf \Psi}}

\def\CC{{\cal C}}

\usepackage{soul}

\newcommand{\ket}[1]{\ensuremath{|#1\rangle}\xspace}
\newcommand{\bra}[1]{\ensuremath{\langle #1|}\xspace}

\begin{document}

\title{Entanglement replication via quantum repeated interactions}

\author{Pierre Wendenbaum}
\affiliation{Institut Jean Lamour, dpt P2M, Groupe de Physique Statistique, Universit\'e de Lorraine-CNRS, B.P. 70239, F-54506 Vand\oe{}uvre-l\`es-Nancy Cedex, France}

\author{Thierry Platini}
\affiliation{Applied Mathematics Research Center, Coventry University, Coventry, England}

\author{Dragi Karevski}
\affiliation{Institut Jean Lamour, dpt P2M, Groupe de Physique Statistique, Universit\'e de Lorraine-CNRS, B.P. 70239, F-54506 Vand\oe{}uvre-l\`es-Nancy Cedex, France}

\date{\today}

\begin{abstract}
We study  entanglement creation between two independent XX chains, which are repeatedly coupled locally to spin-1/2 Bell pairs. We show analytically that in the steady state the entanglement of the Bell pairs is perfectly transferred to the chains, generating large-scale interchain pair correlations. However, before the steady state is reached, within a growing causal region around the interacting locus the chains are found in a current driven nonquilibrium steady state (NESS). In the NESS, the chains cross entanglement decays exponentially with respect to the distance to the boundary sites with a typical length scale which is inversely proportional to the driving current. 
\end{abstract}

\pacs{03.67.Bg, 03.65.Yz, 42.50.Dv}
\maketitle

%
%

Entanglement plays one of the major roles on the scene of quantum engineering and quantum control \cite{Rabitz_2000,Chu_2002}. This purely quantum property is critical in quantum information processing \cite{Bennet_1993,Duan_2001, Reichle_2006}. 
Hence entanglement is the center of attention of numerous studies \cite{Arenz_2013,Schmidt_2013, Zip14,Zippilli_2014_2,Wendenbaum14}.
In a recent work, S. Zippilli and co-workers studied the steady state entanglement replication in
two independent quantum many-body systems, both locally driven by a common entangled field \cite{Zip13,Zip14}. 
They showed that the field can be tuned to perfectly replicate  the driving entanglement across 
the initially independent arrays through the generation of a scale-free set of two-particle Bell states.

In this work, we study the dynamics of entanglement replication between two $XX$-quantum spin chains each of size $N$. The chains are both locally driven at one of their boundaries by an entangled Markovian environment. Starting from any Gaussian initial state, the system reaches, for $t\gg \tau_N$ (with $\tau_N\sim N^3$), a unique stationary state (already observed in \cite{Zip13,Zip14}) in which spins belonging to different chains are perfectly entangled by pairs. During the relaxation toward the stationary state, the system passes through a transient regime for intermediate times $\tau_d \ll t \ll N/v_c$, where $\tau_d$ is a typical microscopic relaxation time related to the local dissipative coupling to the bath and $v_c$ the typical sound velocity within the chains. 
In this time regime, close to the interacting boundaries the chains fall into a Non-Equilibrium Steady State (NESS) with a steady current ${j^z}^*$ injected by the environment into the chains. In the NESS, the entanglement between spins belonging to different chains and facing each other decay exponentially over a length scale $\xi$ as we move away from the interacting boundaries. This entanglement length scale $\xi$ is shown to be proportional to the inverse of the steady current ${j^z}^*$. Moreover, as the NESS extends ballistically through the chains, local observables along the chains show typically a scaling form $Q(x,t)\simeq f_Q(x/t)$ where $x$ is the distance from the interacting boundaries.  

\begin{figure}[h!]
\centering
\includegraphics[width=0.6\columnwidth]{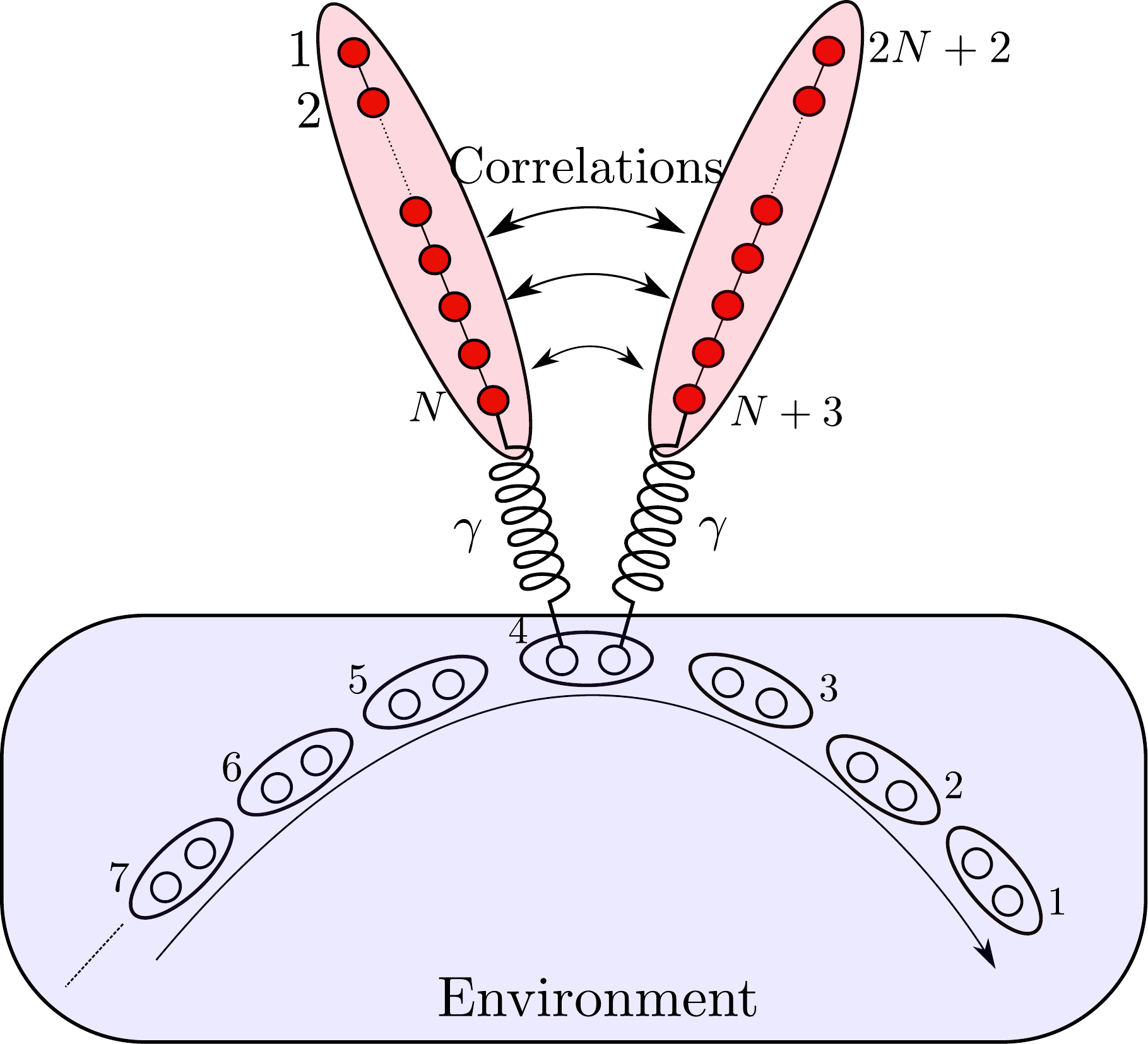}
\caption{(Color online) Two non interacting $XX$ chains are coupled on one edge, via a RIP, to one component of a Bell pair. The first chain is labeled from $1$ to $N$, and the second one from $N+3$ to $2N+2$}
\label{Fig_system}
\end{figure}
%
%

The system  is composed of two non-interacting $XX$ chains with Hamiltonian
\begin{equation}
H_S=-\frac{K}{2}\sum_{n\in \Sbb^\star }\left(\sigma_n^x\sigma_{n+1}^x+\sigma_n^y\sigma_{n+1}^y\right)\;
-\frac{h}{2}\sum_{n\in \Sbb}\sigma_n^z,
\label{hamil1}
\end{equation}
with $\Sbb=\{1,2,\hdots,N\}\cup\{N+3,N+4,\hdots,2N+2\}$ and $\Sbb^\star=\Sbb\backslash\{N,2N+2\}$.  
The hopping constant $K$ is fixed to $1/2$ such that the sound velocity $v_c=1$ into the chains. Note that the total transverse magnetization $M^z=\sum_n\sigma_n^z$ commutes with $H_S$. As a consequence, we will show that the external field does not affect the dynamics of the system, when considering Gaussian initial state.
The Markovian environment with which the system interacts is described by a Repeated Interaction Process (RIP) \cite{Att06,Dha08,KarPla09,PlaHar10,Attal_Deschamps_Pellegrini}, see Fig. \ref{Fig_system} for a pictorial representation. More precisely, the environment is made of an infinite set of identical and independent spin pairs, with Hamiltonians $H_{B}^{(k)}=-{h}(\sigma_{k,1}^z+\sigma_{k,2}^z)/2$ where  $k\in\Nbb^*$, such that the bath Hamiltonian is $H_B=\sum_{k=1}^{\infty}H_{B}^{(k)}$. Each bath spin pair is  prepared into a perfectly entangled Bell state $\eta=\ket{\Phi}\bra{\Phi}$ with $\ket{\Phi}=\frac{1}{\sqrt{2}}\left(\ket{\uparrow\downarrow}+\ket{\downarrow\uparrow}\right)$, where $\ket{\uparrow}$ and $\ket{\downarrow}$ are eigenstates of the $\sigma^z_k$ Pauli matrices. 
Therefore, the initial state of the bath is given by the direct product $\eta_B=\bigotimes_{k\in\Nbb^*}\eta$. 
During the RIP the Bell pairs $\eta_k$ will interact one after the other, over a typical interaction time $\tau$, with the XX-chains.
The dynamics of the full system + environment is generated by the time dependent Hamiltonian $H(t)=H_S+H_B+V(t)$, where $V(t)$ describes the system-environment coupling and is given by $V(t)=V^{(k)}$ for $t\in[(k-1)\tau,k\tau[$.
The interaction couples one spin of the pair, say $\sigma_{k,1}$, to the left chain only while the second one, $\sigma_{k,2}$, is only coupled to the right one and that through a local isotropic $XX$ coupling 
\begin{equation}
V^{(k)}=-\frac{\gamma}{2}\left(\sigma_N^x\sigma_{k,1}^x+\sigma_N^y\sigma_{k,1}^y + \sigma_{k,2}^x\sigma_{N+3}^x+\sigma_{k,2}^y\sigma_{N+3}^y \right)\;. 
\end{equation}
Note that the coupling amplitude $\gamma$ is taken to be the same for the left and right chains. 

The initial  system + environment state is taken to be the tensor product state $\varrho(0)=\varrho_S(0)\otimes \eta_B$ where the system density matrix $\varrho_S(0)$ is a fully factorized (thermal) mixture $\varrho_S(0)=\bigotimes_{n\in\Sbb} \varrho_n$ with the single spin density matrix
$
\varrho_n=(\Id_n+\mu_n\sigma^z_n)/2
$,
where $\mu_n=Tr\{\sigma^z_n \varrho_n \}$. At $\mu_n=1(-1)$ the density matrix $\varrho_n$ reduces to the pure state $| \uparrow \rangle \langle \uparrow |(| \downarrow \rangle \langle \downarrow |)$.

%
%

Over one time step of the RIP, the two chains, together with the Bell pair with which they interact, formally constitute  a $XX$ chain of size $2N+2$ with non-homogeneous couplings.
This chain can be mapped to a free Fermi system through a Jordan Wigner transformation
$c_n=\prod_{j=1}^{n-1}(-\sigma_j^z)\sigma_n^-$ 
 \cite{Lieb1961}
with $2\sigma_n^{\pm}=\sigma^x_n\pm i\sigma^y_n$. The   $c$ and adjoint $c^\dag$ satisfy the canonical anticommutation relations $\{c_i^{\dagger},c_j\}=\delta_{ij}$ and $\{c_i^{\dagger},c_j^{\dagger}\}=\{c_i,c_j\}=0$. To proceed further we separate the system variables from the bath ones. We define the field operator $\Psi^\dag=(\xbf^\dag,\fbf^\dag)$ with $\xbf^{\dagger}=(c_1^{\dagger},\hdots,c_N^{\dag},c_{N+3}^\dag, \hdots, c_{2N+2}^{\dagger})$ and $\fbf^{\dagger}=(c_{N+1}^{\dagger},c_{N+2}^{\dagger})$. The total Hamiltonian takes the form 
$H=\psibf^{\dagger}T \psibf$ where $T$ is a $(2N+2)\times(2N+2)$ matrix given by
\begin{equation}
T=
\begin{pmatrix}
T_S & \gamma\Theta \\
\gamma\Theta^{\dagger} & T_B
\end{pmatrix}-h\mathbb{1} \label{eq_hamil}
, \quad
T_S=
\begin{pmatrix}
A & 0 \\
0 & A
\end{pmatrix},
\end{equation}
where  $A$ (of size $N\times N$) is given by $A_{i,j}=-K(\delta_{i,j+1}+\delta_{i,j-1})$ 
and $\Theta$ (of size $2N\times 2$) is given by $\Theta_{i,j}=-(\delta_{i,N}\delta_{j,1}+\delta_{i,N+1}\delta_{j,2})$.
As long as we are not interested in the fate of the environment Bell pair, it is unnecessary to further specify the $2\times 2$ matrix $T_B$ as it does not appear in the dynamical equation which described the system evolution  \cite{KarPla09}. 

The initial state of the ensemble system+environment being Gaussian \footnote{The initial state of the system+environment is Gaussian and can be written under the form $\varrho\propto\exp(- H_{eff})$ where $H_{eff}$ is quadratic in terms of fermionic operators.}, the  total density matrix $\varrho(t)$ and the system density matrix $\varrho_S(t)=Tr_B\{\varrho(t)\}$ remain Gaussian during the time evolution generated by the free fermionic Hamiltonian \cite{Pes03}. Thanks to Wick theorem, a full description of the system is given in terms of its two points correlators only. 
As a first step we define the $(2N+2)\times(2N+2)$ correlations matrix $G$ by $G_{i,j}(t)=\langle \Psi^{\dagger}_i\Psi_j\rangle(t)$. Its evolution over one time step is given by $G(\tau)=e^{-i\tau T}G(0)e^{i\tau T}$. The magnetic field $h$ appearing on the diagonal of $T$ only, the previous equation clearly indicates that $h$ has no effect on the system dynamics. Next we define the reduced $2N\times 2N$ correlations matrix $G_S$ by $(G_S)_{i,j}(t)=\langle \xbf^{\dagger}_i\xbf_j\rangle(t)$. When taking the continuous time limit ($\tau\rightarrow0$), one has to keep $\gamma^2\tau$ constant \cite{Att06}. This procedure leads to a differential equation describing the evolution of the system in interaction with the environment. The system correlation matrix obeys a Lindblad-like differential equation \cite{KarPla09} 
\begin{eqnarray}\label{EQ_G}
\partial_t G_S=-\mathcal{L}\big(G_S\big)+\Gamma^2\Theta G_B\Theta^{\dagger}\; ,
\end{eqnarray}
with $\mathcal{L}(.)=i[T_S,.]+\Gamma^2\{.,\Theta\Theta^{\dagger}\}/2$ where $\Gamma$ is the rescaled coupling parameter. The $2\times 2$ matrix $G_B$ is the correlation matrix of a bath pair, explicitly given in the Bell state $\eta$ by $(G_B)_{i,j}=1/2$ $\forall i,j$. Choosing the coupling parameter $\Gamma=\gamma\sqrt{\tau}$ leads to the approximate solution for the discrete case. Corrections of order $1/\tau$ are, however expected when considering the discrete case.


%
%

In the following we use the concurrence as a measure of the entanglement between two spins $k,l \in \Sbb$ of the system. For an arbitrary density matrix $\varrho$, representing the quantum state of the two spins, the concurrence \cite{Woo98,ConWoo00} is defined by $\CC=2\max\{0,\lambda_1-\lambda_2-\lambda_3-\lambda_4\}$ where the $\lambda$'s are the square roots of the eigenvalues (in descending order) of the matrix $R=\varrho\tilde{\varrho}$ with $\tilde{\varrho}=(\sigma^y\otimes \sigma^y)\bar{\varrho}(\sigma^y\otimes \sigma^y)$ and where $\bar{\varrho}$ is the complex conjugate of $\varrho$ (in the $\sigma^z$ diagonal base).

%
%
We start, with the simplest case where the chains are reduced to two qubits. In such a case  $T_S=0$, $\Theta=-\mathbb{1}_{2\times2}$ and  Eq. \eqref{EQ_G} becomes $\partial_t G_S=-\Gamma^2(G_S-G_B)$, with solution 
$\langle c_1^{\dagger}c_4\rangle(t) =(1-e^{-\Gamma^2}t)/2$, 
$\langle c_j^{\dagger}c_j\rangle(t)=(1+\mu_je^{-\Gamma^2}t)/2$ for $j=1,4$.
From there, we can reconstruct the reduced density matrix $\varrho^{(1,4)}$
and evaluate the concurrence:
\begin{eqnarray}\label{C_14}
&\CC^{(1,4)}(t)=\max\bigg\{0,{1-e^{-\Gamma^2t}} -\frac{1}{2}\times\\
&\sqrt{\big[2e^{-\Gamma^2t}+(\mu_1\mu_4-1)e^{-2\Gamma^2t}\big]^2
-(\mu_1+\mu_4)^2e^{-2\Gamma^2t}}\bigg\}\;  .\nonumber
\end{eqnarray}
Interestingly enough, one observes a finite delay time $T$, defined as the waiting time needed to generate entanglement [$\CC^{(1,4)}(t)=0$ for all $t<T$], for initially mixed states (for $\mu_n\neq \pm 1$). On the contrary, when the two spins are initially prepared in a pure state then  $T$ vanishes. 
 As an example, in the case of opposite initial magnetizations $\mu_1=-\mu_4=\mu$ one has $T=\Gamma^{-2}\ln\left(1+\sqrt{\frac{1-\mu^2}{2}}\right)$ which vanishes as $(1-|\mu|)^{1/2}$ for $\mu\rightarrow\pm1$.
 This finite waiting time $T$ observed for mixed initial states can be interpreted as the time needed 
 for quantum correlations to overcome the thermal fluctuations, before the entanglement can start to grow more or less with the same functional dependence as in the pure initial state case. 
At long times, independently of the system initial state the concurrence $\CC^{(1,4)}(t\rightarrow\infty)=1$ reflecting the fact that the system correlation matrix $G_S(t\rightarrow \infty)=G_B$. Consequently, in the steady state, the non vanishing system's density matrix coefficients are given by $\varrho^{(1,4)}_{23}=\varrho^{(1,4)}_{32}=-1/2$, $\varrho^{(1,4)}_{22}=\varrho^{(1,4)}_{33}=1/2$ corresponding to the Bell state $\ket{\Phi^-}=\frac{1}{\sqrt{2}}(\ket{\uparrow\downarrow}-\ket{\downarrow\uparrow})$.
In other words, the RIP leads to a transfer of the entanglement from the reservoir to the system. 

%
%
We  consider now  the most interesting case of two chains of equal size $N$. The chains are supposed to be prepared into the same factorized homogeneous state, with initial magnetization $\mu$ on each site. The hopping constant $K$ in (\ref{hamil1}) is fixed to $1/2$ such that the sound velocity $v_c=1$ into the chains.

\begin{figure}[h!]
\centering
\includegraphics[width=0.99\columnwidth]{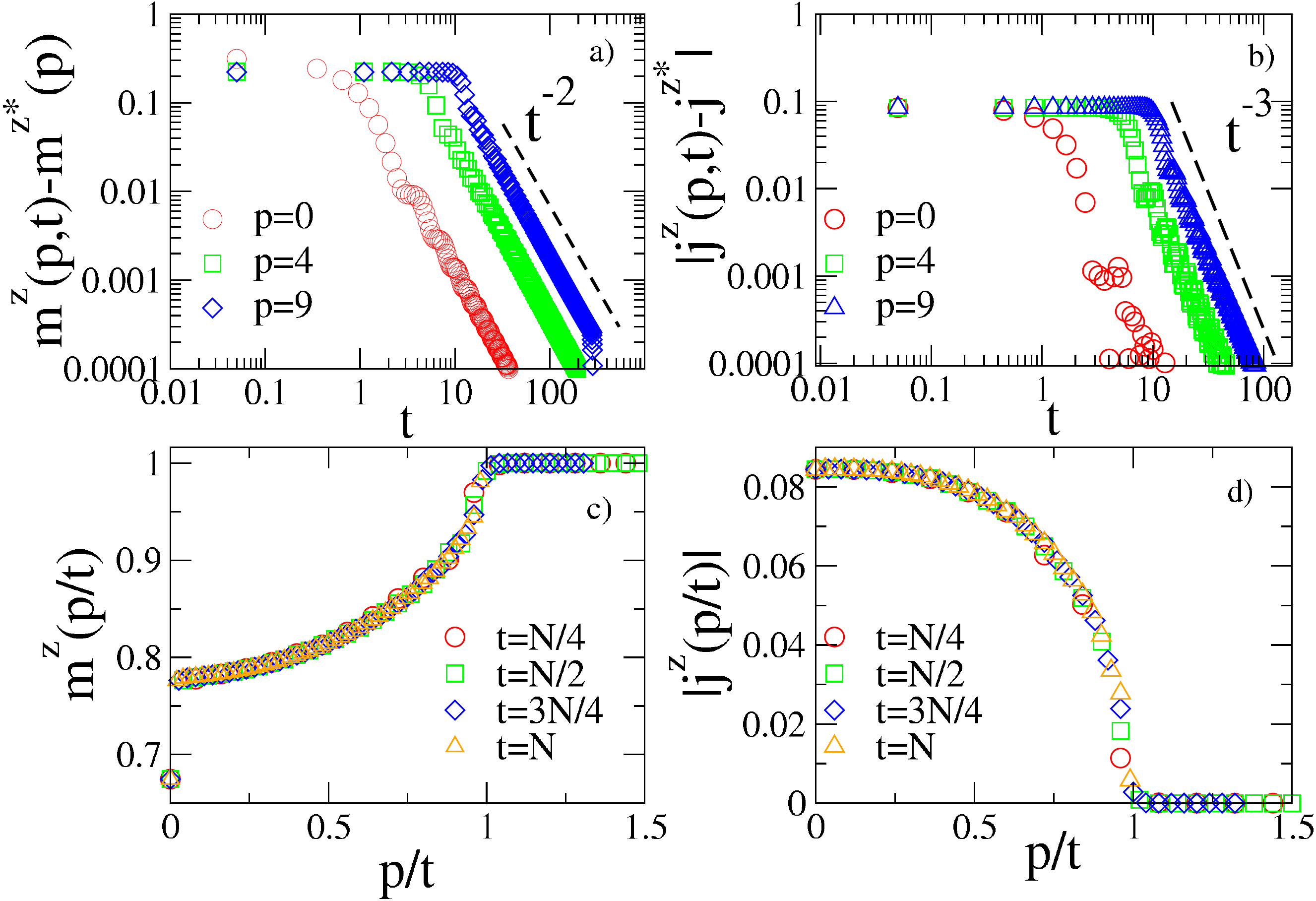}
\caption{(Color online) Time evolution of the magnetization and current for different positions for two chains of size $N=300$, $\Gamma=0.5$ with initial magnetization equal to $1$. 
}
\label{Fig_ness}
\end{figure}

{\it Transient NESS}: 
For sufficiently large system sizes $N$ and times $t<N$, close to the boundaries directly in contact with the environment the system behaves like a semi-infinite one: no excitations have enough time to travel from one boundary to the other. In this regime, the system approaches locally a NESS with a finite flux of excitations injected at the contact point with the environment and traveling into the chains away from the system-environment interaction site.  This flux of particles propagates the new state into the chains, modifying the values of the local quantities and eventually generating correlations. During this process, the transversed magnetization converges algebraically towards a constant value $m^{z*}$: $m^z(p,t)-m^{z*}\propto t^{-2}$, as illustrated in Fig. \ref{Fig_ness}-(a). Outside the immediate vicinity of the boundary point $p=0$ (which corresponds to the sites directly in contact with the bath), the deviation from the asymptotic flat stationary magnetization profile takes the scaling form
$
m^z(p,t)-m^{z*}=f_m(p/t)
$,
as clearly seen on Fig. \ref{Fig_ness}-(c).
The local magnetization current $j^z(p,t)$, which is defined through the continuity equation $\partial_tm^z(x,t)+\partial_xj^z(x,t)=0$ decays as $t^{-3}$ toward an asymptotic value $j^{z*}$ as shown in Fig. \ref{Fig_ness}(b) for different locus $p$ along the chains. Since the current  $j^z(p,t)$ satisfy the continuity equation, it takes also a scaling form $j^z(p,t)=f_j(p/t)$ as seen on Fig. \ref{Fig_ness}-(d).  

\begin{figure}[h!]
\centering
\includegraphics[width=0.99\columnwidth]{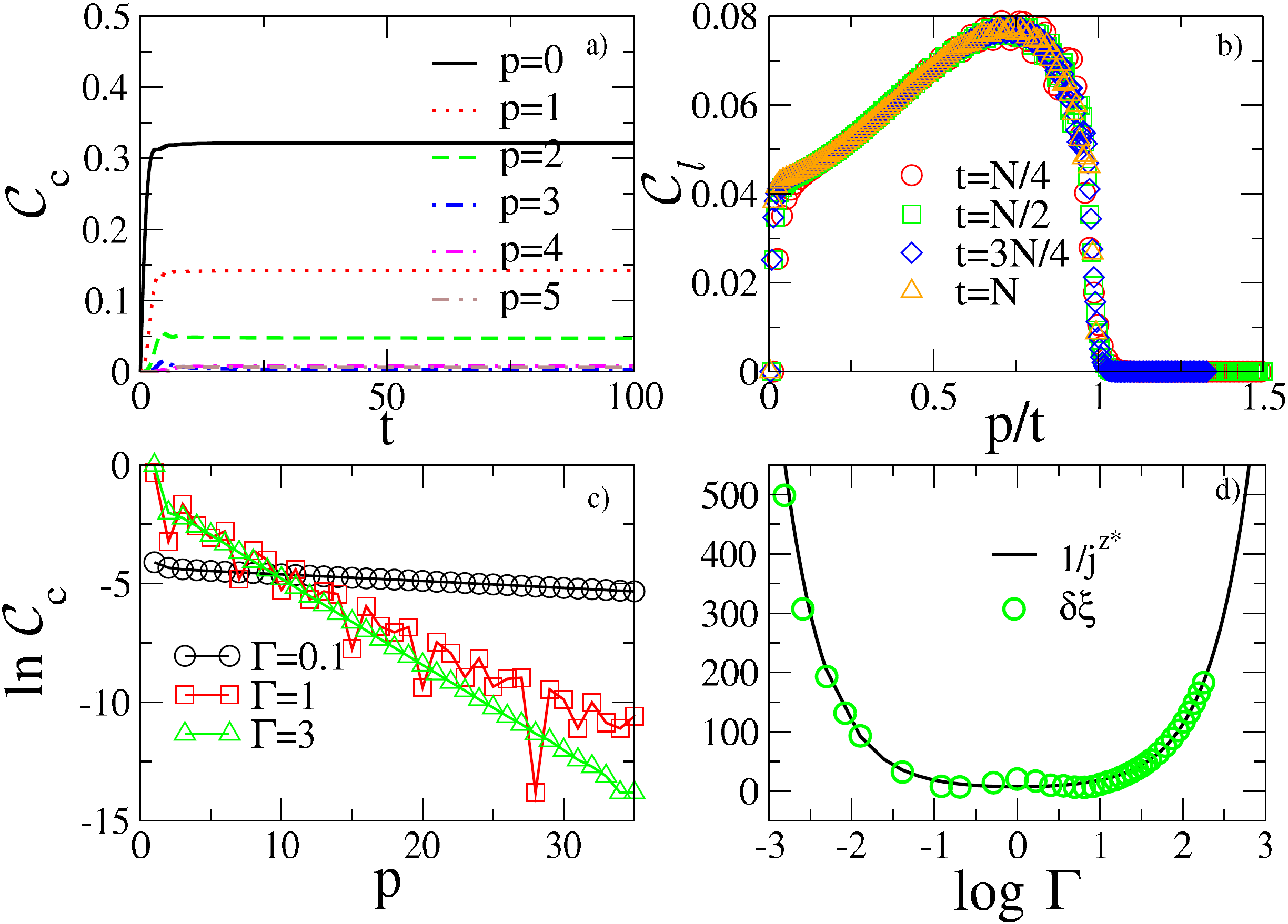}
\caption{(Color online) (a): Concurrence in the pairs $p$ as a function of time. The size of the chains is $N=60$ and the system reservoir coupling is $\Gamma=0.5$. 
(b) Longitudinal concurrence profile at different times as a function of the rescaled parameter $p/t$ of a chain of size $N=500$ and $\Gamma=0.5$. 
(c) Logarithm of the pair concurrence at $t=N$ as a function of the pair $p$ for three different values of $\Gamma$ and $N=60$.
(d) Inverse of the stationary current $j^{z*}$ and entanglement length $\xi$ taken at $t=N$ as a function of the logarithm of the dissipation coupling $\Gamma$ for two chains of size $N=60$. $\beta\simeq 5.67 $ is a proportionality coefficient.  For all plots, the initial magnetization of the chains is $\mu=1$.}
\label{Fig_plateau}
\end{figure}

For $t<N$, the repeated interactions with the  environment generate in the chains steady longitudinal and cross entanglement in the vicinity of the contact point within a region which grows linearly in time. This entanglement is measured respectively by the longitudinal $\CC_{l}(p)$ and cross $\CC_{c}(p)$ concurrences. The later is defined between two spins facing each other (one on site $N-p$ and the other on $N+3+p$ as parametrized in Fig. \ref{Fig_system}). The former, is measured between two neighboring sites $p$ and $p+1$ on a given chain. 
Numerical results obtained from exact diagonalization (see \cite{Ami04} for details) are shown in Fig. \ref{Fig_plateau} for $\mu=1$. 
Mimicking the behavior of the local (in chain) quantities like the magnetization and the current, the longitudinal concurrence $\CC_l(p,t)$ converges toward a steady value by following a scaling form $\CC_l(p,t)=f_{C_l}(p/t)$, which is clearly visible on Fig. \ref{Fig_plateau}-(b). 
Along the repeated interaction process, as already stated cross entanglement is generated between the two chains measured by  $\CC_{c}(p,t)$ and which converges algebraically toward a steady value $\CC_{c}(p)$. 
For $\mu=1$ the steady cross concurrence $\CC_{c}(p)$ decays exponentially with the distance $p$ from the contact point as seen in Figs. \ref{Fig_plateau}-a) and \ref{Fig_plateau}-(c): 
\begin{equation}
C_c(p)\propto \exp\left(-p/\xi\right)\; .
\end{equation}
The decay is over a typical length scale $\xi(\Gamma)$ which seems to be inversely proportional to the steady current $j^{z*}(\Gamma)$ as seen in Fig. \ref{Fig_plateau}-(d). The relation $\xi(\Gamma)\propto 1/j^{z*}(\Gamma)$ is  particularly good for small and large values of $\Gamma$. Deviations can be observed for values around $\Gamma = 1$. Note the symmetry $j^{z*}(\Gamma)=j^{z*}(1/\Gamma)$ as in \cite{KarPla09}. 

\begin{figure}[h!]
\centering
\includegraphics[width=0.99\columnwidth]{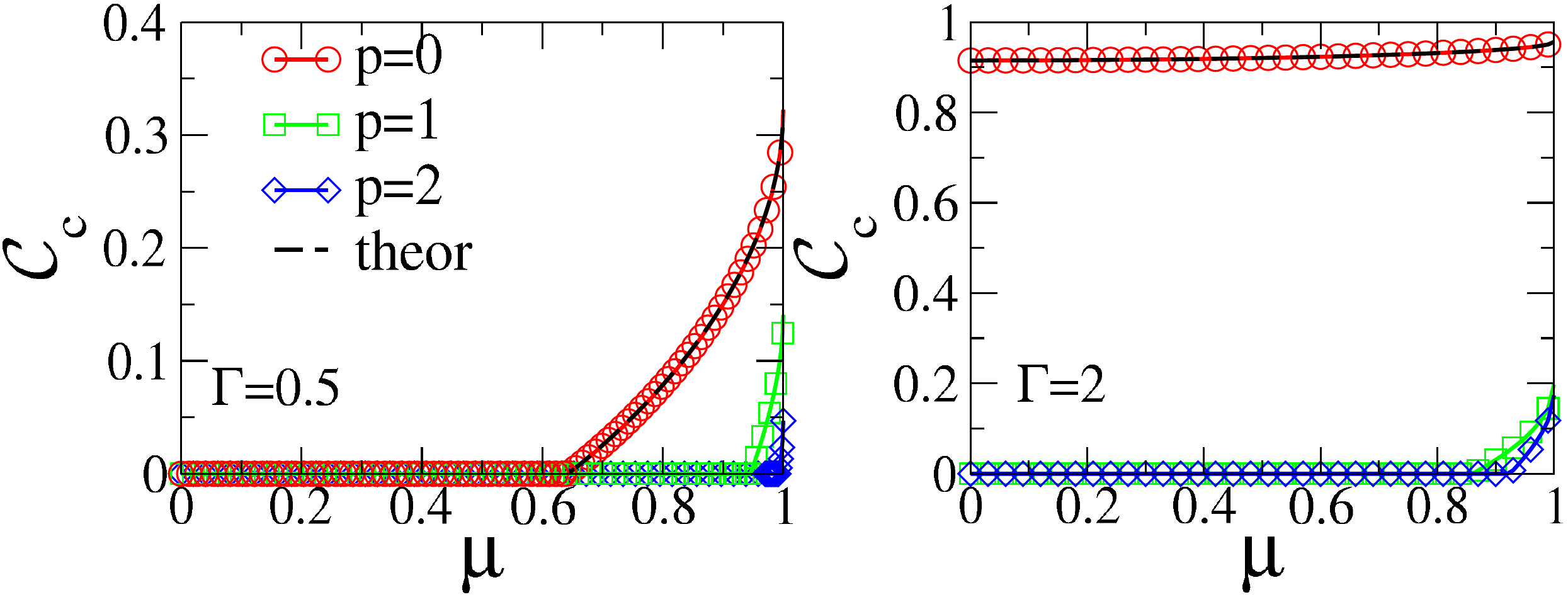}
\caption{(Color online) (a) Cross concurrence in different pairs as a function of the initial magnetization $\mu$ for $\Gamma=1/2$. (b) Same as (a) for $\Gamma=2$. 
The size of the chains is $N=60$. The dots are the numerical results and the dashed line is the theoretical prediction. 
}
\label{Fig_conc_mag}
\end{figure}

For $0<\mu<1$\footnote{We consider here only $\mu>0$ since by symmetry the case $\mu<0$ is trivially deduced.}, the behavior of $\CC_{c}(p)$ is a bit more involved and it is reported in Fig. \ref{Fig_conc_mag} for two different values of $\Gamma$, different positions $p$ and as a function of $\mu$. We see that the cross concurrence decreases, until it reaches $0$, as the initial magnetization $\mu$ decreases.  The value of $\mu$ bellow which a given pair $p$ disentangles converges rapidly toward the saturation value $\mu=1$ as the pair label $p$ is increased. This means that at finite initial temperature ($\mu<1$) very few pairs are entangled in the NESS. For $\Gamma=1/2$ we also see that there is a threshold value $\mu_{thre}$ bellow which there is no more cross entanglement. On the contrary at larger dissipation rate $\Gamma$,  for example $\Gamma=2$, the first pair is always entangled to a very high value (close to one) whatever the initial magnetization is, the price to be paid being that the remaining pairs are already disentangled for values of $\mu<0.9$. 

This intriguing behavior of the cross concurrence can be explained in the following way (at least for the first pair $p=0$): assuming that the magnetization of spins N and N + 3 in the NESS is proportional to the initial magnetization of the chain, $m_{N,N+3}=(1-\alpha)\mu$ and that the fermionic correlator is proportional to the Bell pairs correlations, 
$\langle c_N^\dagger c_{N+3}\rangle= \alpha  \langle c_{N+1}^\dagger c_{N+2}\rangle=\alpha/2$, 
where the coefficient $\alpha$ is a function of $\Gamma$. 
It follows by solving locally the steady Lindblad equations that the cross concurrence of the first pair $p=0$ is given by
\begin{equation}
C_c(0)= \max \left\{ 0, \alpha -\frac{1}{2}\sqrt{g(\alpha,\mu)}
\right\}
\label{Cc_analytic}
\end{equation}
with $g(\alpha,\mu)=(1-\alpha)^2(1-\mu^2)[(1+\alpha)^2-\mu^2(1-\alpha)^2]$. For $\mu= 1$ one has $C_c(0)=\alpha$ which fixes then the value of $\alpha$ for a given $\Gamma$. Depending on the value of $\alpha$ there may or may not exist a threshold value of $\mu$, $\mu_{thre}$ defined by $C_c(0)=0\; \forall \mu<\mu_{thre}$, below which the entanglement in the first pair is lost in the NESS. For $\alpha> \sqrt{2}-1$ (which corresponds to $\Gamma\simeq 0.5916$), the first pair is always entangled in the NESS whatever the value of $\mu$. In Fig. \ref{Fig_conc_mag} the theoretical prediction (\ref{Cc_analytic}) is plotted for the two different values of $\Gamma$ and one can see that it matches perfectly the numerical results. 

After the NESS regime, the system undergoes a relaxation regime which results from the superposition of constructive and destructive interferences of excitations traveling up and down the chains, being reflected by the boundaries many times. This process leads to an exponential relaxation towards the stationary state with a relaxation time diverging with the system size ($\tau\propto N^3$). In the following we focus on the stationary state reached in the limit $t\rightarrow\infty$ and satisfying $\mathcal{L}\big(G_S^*\big)=\Gamma^2\Theta G_B\Theta^{\dagger}$.

{\it Steady state}. 
At very large times, due to multiple reflections  at the chains boundaries of the excitations injected by the bath, the local NESS  is progressively lost. The system converges toward its real unique stationary state with a typical relaxation time of order $N^3$. This final stationary state can be obtained in the following way:
Split the correlation function $G_{S,B}$ into a real and imaginary part, $G_k=N_k+iJ_k$ $(k=S,B)$ with $N_k^T=N_k$ and $J_k^T=-J_k$. The $J_S$ matrix contains current-like terms of the form $i\langle c^\dag_nc_m-c^\dag_mc_n\rangle$ which have to vanish in the steady state and the remaining matrix $N_S$ has to satisfy from (\ref{EQ_G})  the steady equations 
$
[T_S,N_S]=0\; , \
\{\Theta\Theta^{\dagger},N_S\}=2\Theta N_B\Theta^{\dagger}
$.
The only non-vanishing entries of the solution matrix $N_S$ are  the on-site densities 
$\langle c_i^{\dagger}c_i\rangle=1/2$  and the cross-correlations 
$\langle c_i^{\dagger}c_{2N+3-i}\rangle=1/2$, $\forall i \in\Sbb$. It follows that only sites facing each other are correlated and the steady state turns out to be given by an alternation of pairs in the perfectly entangled states $\ket{\Phi^{\pm}}=\frac{1}{\sqrt{2}}(\ket{\uparrow\downarrow}\pm\ket{\downarrow\uparrow})$.
The steady density matrix factorizes into 
$
\varrho_S^*=\otimes_{p}\varrho_p
$
with $\varrho_{2n}=\ket{\Phi^{-}}\bra{\Phi^{-}}$ and $\varrho_{2n+1}=\ket{\Phi^{+}}\bra{\Phi^{+}}$.

In summary the replication mechanism is perfect in the steady state.  However, and most interestingly, in an intermediate NESS regime, for pure initial state ($|\mu|=1$), the cross entanglement between the chains decays exponentially with respect to the distance to the interacting locus. This exponential decay is over a typical length scale which is found to be inversely proportional to the driving current. At finite temperatures ($|\mu|<1$) and at low enough dissipation rate, there is a threshold magnetization below which the cross entanglement between the chains is completely lost in the NESS.

\bibliographystyle{apsrev}
\bibliography{biblio}

\begin{thebibliography}{21}
\expandafter\ifx\csname natexlab\endcsname\relax\def\natexlab#1{#1}\fi
\expandafter\ifx\csname bibnamefont\endcsname\relax
  \def\bibnamefont#1{#1}\fi
\expandafter\ifx\csname bibfnamefont\endcsname\relax
  \def\bibfnamefont#1{#1}\fi
\expandafter\ifx\csname citenamefont\endcsname\relax
  \def\citenamefont#1{#1}\fi
\expandafter\ifx\csname url\endcsname\relax
  \def\url#1{\texttt{#1}}\fi
\expandafter\ifx\csname urlprefix\endcsname\relax\def\urlprefix{URL }\fi
\providecommand{\bibinfo}[2]{#2}
\providecommand{\eprint}[2][]{\url{#2}}

\bibitem[{\citenamefont{Rabitz et~al.}(2000)\citenamefont{Rabitz,
  de~Vivie-Riedle, Motzkus, and Kompa}}]{Rabitz_2000}
\bibinfo{author}{\bibfnamefont{H.}~\bibnamefont{Rabitz}},
  \bibinfo{author}{\bibfnamefont{R.}~\bibnamefont{de~Vivie-Riedle}},
  \bibinfo{author}{\bibfnamefont{M.}~\bibnamefont{Motzkus}}, \bibnamefont{and}
  \bibinfo{author}{\bibfnamefont{K.}~\bibnamefont{Kompa}},
  \bibinfo{journal}{Science} \textbf{\bibinfo{volume}{288}},
  \bibinfo{pages}{824} (\bibinfo{year}{2000}).

\bibitem[{\citenamefont{Chu}(2002)}]{Chu_2002}
\bibinfo{author}{\bibfnamefont{S.}~\bibnamefont{Chu}},
  \bibinfo{journal}{Nature (London)} \textbf{\bibinfo{volume}{416}},
  \bibinfo{pages}{206} (\bibinfo{year}{2002}).

\bibitem[{\citenamefont{Bennett et~al.}(1993)\citenamefont{Bennett, Brassard,
  Cr\'epeau, Jozsa, Peres, and Wootters}}]{Bennet_1993}
\bibinfo{author}{\bibfnamefont{C.}~\bibnamefont{Bennett}},
  \bibinfo{author}{\bibfnamefont{G.}~\bibnamefont{Brassard}},
  \bibinfo{author}{\bibfnamefont{C.}~\bibnamefont{Cr\'epeau}},
  \bibinfo{author}{\bibfnamefont{R.}~\bibnamefont{Jozsa}},
  \bibinfo{author}{\bibfnamefont{A.}~\bibnamefont{Peres}}, \bibnamefont{and}
  \bibinfo{author}{\bibfnamefont{W.~K.} \bibnamefont{Wootters}},
  \bibinfo{journal}{Phys. Rev. Lett.} \textbf{\bibinfo{volume}{70}},
  \bibinfo{pages}{1895} (\bibinfo{year}{1993}).

\bibitem[{\citenamefont{Duan et~al.}(2001)\citenamefont{Duan, Lukin, Cirac, and
  Zoller}}]{Duan_2001}
\bibinfo{author}{\bibfnamefont{L.~M.} \bibnamefont{Duan}},
  \bibinfo{author}{\bibfnamefont{M.~D.} \bibnamefont{Lukin}},
  \bibinfo{author}{\bibfnamefont{J.~I.} \bibnamefont{Cirac}}, \bibnamefont{and}
  \bibinfo{author}{\bibfnamefont{P.}~\bibnamefont{Zoller}},
  \bibinfo{journal}{Nature (London)} \textbf{\bibinfo{volume}{414}},
  \bibinfo{pages}{413} (\bibinfo{year}{2001}).

\bibitem[{\citenamefont{Reichle et~al.}(2006)\citenamefont{Reichle, Leibfried,
  Knill, Britton, Blakestad, Jost, Langer, Ozeri, Seidelin, and
  Wineland}}]{Reichle_2006}
\bibinfo{author}{\bibfnamefont{R.}~\bibnamefont{Reichle}},
  \bibinfo{author}{\bibfnamefont{D.}~\bibnamefont{Leibfried}},
  \bibinfo{author}{\bibfnamefont{E.}~\bibnamefont{Knill}},
  \bibinfo{author}{\bibfnamefont{J.}~\bibnamefont{Britton}},
  \bibinfo{author}{\bibfnamefont{R.~B.} \bibnamefont{Blakestad}},
  \bibinfo{author}{\bibfnamefont{J.~D.} \bibnamefont{Jost}},
  \bibinfo{author}{\bibfnamefont{C.}~\bibnamefont{Langer}},
  \bibinfo{author}{\bibfnamefont{R.}~\bibnamefont{Ozeri}},
  \bibinfo{author}{\bibfnamefont{S.}~\bibnamefont{Seidelin}}, \bibnamefont{and}
  \bibinfo{author}{\bibfnamefont{D.~J.} \bibnamefont{Wineland}},
  \bibinfo{journal}{Nature (London)} \textbf{\bibinfo{volume}{443}},
  \bibinfo{pages}{838} (\bibinfo{year}{2006}).

\bibitem[{\citenamefont{Arenz et~al.}(2013)\citenamefont{Arenz, Cormick,
  Vitali, and Morigi}}]{Arenz_2013}
\bibinfo{author}{\bibfnamefont{C.}~\bibnamefont{Arenz}},
  \bibinfo{author}{\bibfnamefont{C.}~\bibnamefont{Cormick}},
  \bibinfo{author}{\bibfnamefont{D.}~\bibnamefont{Vitali}}, \bibnamefont{and}
  \bibinfo{author}{\bibfnamefont{G.}~\bibnamefont{Morigi}},
  \bibinfo{journal}{J. Phys. B: At. Mol. Opt. Phys.}
  \textbf{\bibinfo{volume}{46}}, \bibinfo{pages}{224001}
  (\bibinfo{year}{2013}).

\bibitem[{\citenamefont{Schmidt et~al.}(2013)\citenamefont{Schmidt,
  Stockburger, and Ankerhold}}]{Schmidt_2013}
\bibinfo{author}{\bibfnamefont{R.}~\bibnamefont{Schmidt}},
  \bibinfo{author}{\bibfnamefont{J.~T.} \bibnamefont{Stockburger}},
  \bibnamefont{and}
  \bibinfo{author}{\bibfnamefont{J.}~\bibnamefont{Ankerhold}},
  \bibinfo{journal}{Phys. Rev. A} \textbf{\bibinfo{volume}{88}},
  \bibinfo{pages}{052321} (\bibinfo{year}{2013}).

\bibitem[{\citenamefont{Zippilli and Illuminati}(2014)}]{Zip14}
\bibinfo{author}{\bibfnamefont{S.}~\bibnamefont{Zippilli}} \bibnamefont{and}
  \bibinfo{author}{\bibfnamefont{F.}~\bibnamefont{Illuminati}},
  \bibinfo{journal}{Phys. Rev. A} \textbf{\bibinfo{volume}{89}},
  \bibinfo{pages}{033803} (\bibinfo{year}{2014}).

\bibitem[{\citenamefont{Zippilli et~al.}(2014)\citenamefont{Zippilli, Grajcar,
  Il'ichev, and Illuminati}}]{Zippilli_2014_2}
\bibinfo{author}{\bibfnamefont{S.}~\bibnamefont{Zippilli}},
  \bibinfo{author}{\bibfnamefont{M.}~\bibnamefont{Grajcar}},
  \bibinfo{author}{\bibfnamefont{E.}~\bibnamefont{Il'ichev}}, \bibnamefont{and}
  \bibinfo{author}{\bibfnamefont{F.}~\bibnamefont{Illuminati}},
  \bibinfo{journal}{arXiv:1410.5444}  (\bibinfo{year}{2014}).

\bibitem[{\citenamefont{Wendenbaum et~al.}(2014)\citenamefont{Wendenbaum,
  Taketani, and Karevski}}]{Wendenbaum14}
\bibinfo{author}{\bibfnamefont{P.}~\bibnamefont{Wendenbaum}},
  \bibinfo{author}{\bibfnamefont{B.}~\bibnamefont{Taketani}}, \bibnamefont{and}
  \bibinfo{author}{\bibfnamefont{D.}~\bibnamefont{Karevski}},
  \bibinfo{journal}{Phys. Rev. A} \textbf{\bibinfo{volume}{90}},
  \bibinfo{pages}{022125} (\bibinfo{year}{2014}).

\bibitem[{\citenamefont{Zippilli et~al.}(2013)\citenamefont{Zippilli,
  Paternostro, Adesso, and Illuminati}}]{Zip13}
\bibinfo{author}{\bibfnamefont{S.}~\bibnamefont{Zippilli}},
  \bibinfo{author}{\bibfnamefont{M.}~\bibnamefont{Paternostro}},
  \bibinfo{author}{\bibfnamefont{G.}~\bibnamefont{Adesso}}, \bibnamefont{and}
  \bibinfo{author}{\bibfnamefont{F.}~\bibnamefont{Illuminati}},
  \bibinfo{journal}{Phys. Rev. Lett.} \textbf{\bibinfo{volume}{110}},
  \bibinfo{pages}{040503} (\bibinfo{year}{2013}).

\bibitem[{\citenamefont{Attal and Pautrat}(2006)}]{Att06}
\bibinfo{author}{\bibfnamefont{S.}~\bibnamefont{Attal}} \bibnamefont{and}
  \bibinfo{author}{\bibfnamefont{Y.}~\bibnamefont{Pautrat}},
  \bibinfo{journal}{Ann. Henri Poincar\'e} \textbf{\bibinfo{volume}{7}},
  \bibinfo{pages}{59} (\bibinfo{year}{2006}).

\bibitem[{\citenamefont{Dhahri}(2008)}]{Dha08}
\bibinfo{author}{\bibfnamefont{A.}~\bibnamefont{Dhahri}}, \bibinfo{journal}{J.
  Phys. A: Math. Theor.} \textbf{\bibinfo{volume}{41}}, \bibinfo{pages}{275305}
  (\bibinfo{year}{2008}).

\bibitem[{\citenamefont{Karevski and Platini}(2009)}]{KarPla09}
\bibinfo{author}{\bibfnamefont{D.}~\bibnamefont{Karevski}} \bibnamefont{and}
  \bibinfo{author}{\bibfnamefont{T.}~\bibnamefont{Platini}},
  \bibinfo{journal}{Phys. Rev. Lett.} \textbf{\bibinfo{volume}{102}},
  \bibinfo{pages}{207207} (\bibinfo{year}{2009}).

\bibitem[{\citenamefont{Platini et~al.}(2010)\citenamefont{Platini, Harris, and
  Karevski}}]{PlaHar10}
\bibinfo{author}{\bibfnamefont{T.}~\bibnamefont{Platini}},
  \bibinfo{author}{\bibfnamefont{R.~J.} \bibnamefont{Harris}},
  \bibnamefont{and} \bibinfo{author}{\bibfnamefont{D.}~\bibnamefont{Karevski}},
  \bibinfo{journal}{J. Phys. A: Math. Theor.} \textbf{\bibinfo{volume}{43}},
  \bibinfo{pages}{135003} (\bibinfo{year}{2010}).

\bibitem[{\citenamefont{Attal et~al.}(2014)\citenamefont{Attal, Deschamps, and
  Pellegrini}}]{Attal_Deschamps_Pellegrini}
\bibinfo{author}{\bibfnamefont{S.}~\bibnamefont{Attal}},
  \bibinfo{author}{\bibfnamefont{J.}~\bibnamefont{Deschamps}},
  \bibnamefont{and}
  \bibinfo{author}{\bibfnamefont{C.}~\bibnamefont{Pellegrini}},
  \bibinfo{journal}{Journal of Statistical Physics}
  \textbf{\bibinfo{volume}{154}}, \bibinfo{pages}{819} (\bibinfo{year}{2014}).

\bibitem[{\citenamefont{Lieb et~al.}(1961)\citenamefont{Lieb, Schultz, and
  Mattis}}]{Lieb1961}
\bibinfo{author}{\bibfnamefont{E.}~\bibnamefont{Lieb}},
  \bibinfo{author}{\bibfnamefont{T.}~\bibnamefont{Schultz}}, \bibnamefont{and}
  \bibinfo{author}{\bibfnamefont{D.}~\bibnamefont{Mattis}},
  \bibinfo{journal}{Ann. Phys. (NY)} \textbf{\bibinfo{volume}{16}},
  \bibinfo{pages}{407} (\bibinfo{year}{1961}).

\bibitem[{\citenamefont{Peschel}(2003)}]{Pes03}
\bibinfo{author}{\bibfnamefont{I.}~\bibnamefont{Peschel}}, \bibinfo{journal}{J.
  Phys. A: Math. and General} \textbf{\bibinfo{volume}{36}},
  \bibinfo{pages}{L205} (\bibinfo{year}{2003}).

\bibitem[{\citenamefont{Wootters}(1998)}]{Woo98}
\bibinfo{author}{\bibfnamefont{W.~K.} \bibnamefont{Wootters}},
  \bibinfo{journal}{Phys. Rev. Lett.} \textbf{\bibinfo{volume}{80}},
  \bibinfo{pages}{2245} (\bibinfo{year}{1998}).

\bibitem[{\citenamefont{O'Connor and Wootters}(2001)}]{ConWoo00}
\bibinfo{author}{\bibfnamefont{K.~M.} \bibnamefont{O'Connor}} \bibnamefont{and}
  \bibinfo{author}{\bibfnamefont{W.~K.} \bibnamefont{Wootters}},
  \bibinfo{journal}{Phys. Rev. A} \textbf{\bibinfo{volume}{63}},
  \bibinfo{pages}{052302} (\bibinfo{year}{2001}).

\bibitem[{\citenamefont{Amico et~al.}(2004)\citenamefont{Amico, Osterloh,
  Plastina, Fazio, and Palma}}]{Ami04}
\bibinfo{author}{\bibfnamefont{L.}~\bibnamefont{Amico}},
  \bibinfo{author}{\bibfnamefont{A.}~\bibnamefont{Osterloh}},
  \bibinfo{author}{\bibfnamefont{F.}~\bibnamefont{Plastina}},
  \bibinfo{author}{\bibfnamefont{R.}~\bibnamefont{Fazio}}, \bibnamefont{and}
  \bibinfo{author}{\bibfnamefont{G.~M.} \bibnamefont{Palma}},
  \bibinfo{journal}{Phys. Rev. A} \textbf{\bibinfo{volume}{69}},
  \bibinfo{pages}{022304} (\bibinfo{year}{2004}).

\end{thebibliography}





\end{document}